\documentclass[twocolumn,showpacs,preprintnumbers,amsmath,amssymb,pra]{revtex4}

\usepackage{graphicx}
\usepackage{dcolumn}
\usepackage{bm}
\usepackage{epsfig}
\newcommand{\be}{\begin{eqnarray}}
\newcommand{\ee}{\end{eqnarray}}

\begin{document}

\title{Universality class of quantum criticality for strongly repulsive spin-1 bosons with antiferromagnetic spin-exchange interaction}

\author{C. C. N. Kuhn$^{1,2}$, X. W. Guan$^{2}$\footnote{xwe105@physics.anu.edu.au}, A. Foerster$^{1}$ and M. T. Batchelor$^{2,3}$ }

\affiliation{$^{1}$ Instituto de Fisica da UFRGS, Av. Bento Goncalves 9500, Porto Alegre, RS, Brazil\\
$^{2}$ Department of Theoretical Physics, Research School of Physics and
Engineering, Australian National University, Canberra ACT 0200, Australia\\
$^{3}$ Mathematical Sciences Institute, Australian National University, Canberra ACT 0200, Australia
}

\date{\today}

\begin{abstract}
Using the thermodynamic Bethe ansatz equations we study the quantum phase diagram,  thermodynamics  and criticality  of one-dimensional 
spin-1 bosons with strongly repulsive density-density and antiferromagnetic spin-exchange interactions.
We analytically derive  a  high precision equation of state 
from which the Tomonaga-Luttinger liquid physics and quantum critical behavior  of the system  are computed. We obtain explicit forms for   the scaling functions near the critical points yielding 
the dynamical exponent $z=2$ and correlation length exponent $\nu=1/2$  for the  quantum  phase transitions driven 
by either the chemical potential or the  magnetic field.  Consequently, we further demonstrate that quantum criticality of the system can be mapped out from the finite temperature  density and magnetization  profiles of the 1D  trapped gas. 
Our  results  provide the physical  origin of quantum criticality in a 1D many-body system beyond the Tomonaga-Luttinger liquid description.   

\end{abstract}



\maketitle
\section{INTRODUCTION}
The study of spinor Bose  gases  is an active area of research in the field of cold atoms   \cite{Ho,Ohmi}.  In an  optical trap,  the laser-atom 
interaction is determined by the induced electric dipole moment, thus  the atoms are confined independently of their spin orientations. This has provided 
exciting opportunities of simulating quantum dynamics of  spinor Bose-Einstein condensates in which  the ``vector" property of spinor atoms can be  preserved.  Several experimental 
groups have successfully demonstrated spinor BECs of   $^{23}$Na \cite{Na1,Na2}  and $^{87}$Rb \cite{matt,barr,Chang}  atoms in optical traps.  In particular,  the exquisite  
tunability with ultracold atoms confined to  low dimensions  has  provided unprecedented opportunities   for testing the theory of one-dimensional (1D) exactly solvable many-body 
systems \cite{Moritz,Paredes,Exp1,Exp2,Exp3,Exp4,Kitagawa,Armijo,Exp7,Liao}.  These experimental  developments  have stimulated  an  extensive  study of related exactly solvable models 
with $\delta$-function interactions, see  recent  reviews \cite{1D-1,Cazalilla}

\begin{figure}[t]
{{\includegraphics [width=0.82\linewidth,angle=-90]{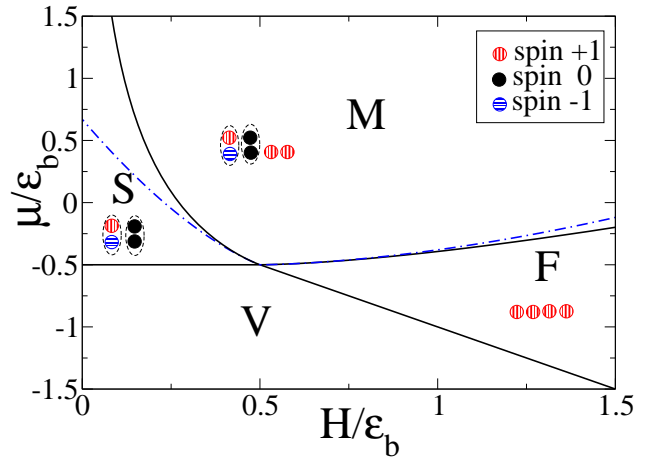}}}
\caption{(Color online) Phase diagram in the $\mu$-$H$ plane showing the    spin-singlet phase $S$ of paired bosons,  
ferromagnetic phase $F$ of spin-aligned bosons, and  a mixed phase $M$ 
of  pairs  and unpaired bosons.  $V$ stands for  the vacuum. The $S-M$ and $F-M$ boundaries are determined by   the critical fields  (\ref{mu3_ep}) and (\ref{mu4_ep}), respectively.  
The $V-S$ and $V-F$ boundaries are given in   (\ref{mu_exac1}).  The dashed-dotted  lines are the extrapolation of the phase boundaries  (\ref{mu3_ap}) and (\ref{mu4_ap}) in the strong 
coupling regime.}
\label{fig1} 
\end{figure}

Quantum spinor  gases with multi-spin states  exhibit  richer  quantum effects than their single component counterparts.  Spinor Bose gases with spin-independent 
short range interaction have a ferromagnetic ground state, i.e. the ground state is always fully polarized \cite{Eisenberg-Lieb,Yang-Li}.   
In contrast to the two-component Fermi gases \cite{Yang-Gaudin}, the two-component  spinor Bose gas with spin-independent s-wave scattering  \cite{Sutherland,Li,GBT} 
has a ferromagnetic ground  state as long as the interaction is fully spin independent.  However, 1D  spinor Bose gas  with  short-range density-density and  
spin-exchange  interactions \cite{Cao,Cao2} can display a different ground state, i.e., either  a ferromagnetic or an antiferromagnetic ground state  solely 
depending on the spin-exchange  interaction.  In this context,  the spin-1 spinor Bose gas with  short-range delta-function  interaction and antiferromagnetic 
spin-spin interaction  is particularly interesting due to the existence of  various phases of quantum liquids associated with the 
 Bethe Ansatz (BA) solutions \cite{Cao,Shlyapnikov,Lee,Shlyapnikov2}.  At zero temperature, this  model  exhibits three 
phases   in the chemical potential -- magnetic field plane. These are (i) a spin-singlet phase of pairs of  bosons with hyperfine states $|F = 1,m_F=\pm 1 \rangle$ 
or two $|F = 1, m_F=0\rangle $ bosons, (ii) a ferromagnetic 
phase of fully-polarized atoms  in the hyperfine state $|F = 1,m_F = 1\rangle$  and  (iii) a mixed phase of spin-singlet pairs and unpaired single atoms, see Figure \ref{fig1}.

Spinor Bose gases  exhibit  various phases of strongly correlated quantum liquids and are 
thus particularly valuable to investigate
 quantum magnetism and criticality. Near a quantum critical point, the many-body system is expected to show universal scaling 
behaviour in the thermodynamic quantities due to the collective nature of many-body effects. Thus  a universal and scale-invariant description of the system is 
expected through the power-law scaling of thermodynamic properties \cite{sachdev,Fisher}. Most recently,  quantum criticality and universal scaling behaviour have 
been experimentally investigated  in low-dimensional cold atomic matter \cite{Exp10,Exp11}. These advances build on  theoretical schemes for mapping out quantum 
criticality in cold atom systems \cite{Campostrini, Zhou-Ho,Erich}. In this framework, exactly solvable models of cold atoms, exhibiting 
quantum phase transitions,  provide  a rigorous way to  treat  quantum criticality in   archetypical quantum many-body systems, such as the 
Gaudin-Yang Fermi gas  \cite{Guan-Ho}, the Lieb-Liniger Bose gas \cite{GB} and a  mixture of bosons and fermions \cite{Yin-G}.

Despite much work on the 1D  spin-1 bosons with strongly repulsive density-density and antiferromagnetic spin-exchange interactions  \cite{Cao,Shlyapnikov,Lee}, there 
has been no  study of the quantum criticality of the model  by using the exact solution. 
Contrary to the impression one might have, exact solvability does not guarantee that physical quantities of
interest can be actually calculated by the BA solutions. The thermodynamic
Bethe ansatz (TBA) equations for this model \cite{Lee}  involve an infinite number of coupled nonlinear
integral equations that impose a number of challenges to access the physics of the model. 

 In the present paper, building on the method proposed in the study of quantum criticality of the Gaudin-Yang Fermi gas \cite{Guan-Ho} and  the 
 Lieb-Liniger Bose gas \cite{GB}, we analytically 
study the  quantum phase diagram,  universal  thermodynamics and criticality of 
spin-1 bosons with strongly repulsive density-density and antiferromagnetic spin-exchange interactions.
We derive a high precision  equation of state of the system in  experimental accessible conditions, i.e.,  in the strong coupling regime and low temperatures. We also 
analytically derive the Tomonaga-Luttinger liquid (TLL)  thermodynamics, quantum critical exponents   and universal scaling functions near the critical points 
associated with quantum  phase transitions driven by the  chemical potential and magnetic field. 
These scaling forms for the thermodynamic  properties across the phase boundaries 
 illustrate the physical origin of quantum criticality in this system, where the singular part of the 
 thermodynamic properties involves a sudden change of density of state for either pairs or unpaired single atoms.  

The paper is organized as follows. In Sec \ref{Model},  we present the model and its corresponding TBA equations.
In Sec \ref{Phase}, we  analytically determine the phase diagram of the model  at zero temperature. 
In Sec \ref{EOS},  we derive the equation of state and universal TLL thermodynamics in the physical regime, i.e., for strong coupling and low temperatures.  
In Sec \ref{QC}, we investigate quantum critical behaviour driven by the chemical potential and magnetic field. The scaling functions near  the critical 
points are obtained analytically.  Sec \ref{conclusion} is  the conclusion.

\section{THE MODEL}
\label{Model}

 We consider $N$ particles of mass $m$ confined in 1D to a length $L$ with 
$\delta$-interacting type density-density and spin-exchange interactions between two atoms.  The 
Hamiltonian is given by \cite{Ho,Cao}
\begin{eqnarray}
 {\cal H}=-\sum^N_{i=1}\frac{\partial^2}{\partial x^2_i}+\sum_{i<j}[c_0+c_2S_i\cdot S_j]\delta(x_i-x_j)+E_z,
\label{Ham}
\end{eqnarray}
where  $S_i$ is the spin-1 operator with $z$-component $(s=1, 0, -1)$. The interaction parameters  $c_0=(g_0+2g_2)/3$ and $c_2=(g_2-g_0)/3$ where 
$g_S=4\pi\hbar^2a_S/m$. Here $m$ is the particle mass and $a_S$ represents the $s$-wave scattering length in the total spin $S=0,2$ channels. 
$E_z=-HS^z$ stands for the Zeeman energy, where $H$ is the external field and $S^z$ the total spin in the 
$z$-component. In the above equation, we have set  $\hbar=2m=1$. 

Using the BA hypothesis, Cao {\em et al.} \cite{Cao}
solved the model  (\ref{Ham})  with antiferromagnetic spin-exchange interaction for $c=c_0=c_2>0$. 
The energy eigenspectrum is given in terms of  the quasi-momenta $\{k_j\}$ of the particles through $E= \sum_{j=1}^{N}k_j^2$, obeying 
the following set of coupled BA equations \cite{Cao}
\begin{eqnarray}
&& \exp(\mathrm{i}k_jL) = \prod^N_{\ell=1}\frac{k_j-k_{\ell}+4c'\mathrm{i}}{k_j-k_{\ell}-4c'\mathrm{i}}\prod^M_{\alpha = 1}\frac{k_j-\Lambda_\alpha-2c'\mathrm{i}}{k_j-\Lambda_\alpha+2c'\mathrm{i}},\nonumber\\
&& \prod^N_{\ell = 1}\frac{\Lambda_{\alpha}-k_{\ell}+2c'\mathrm{i}}{\Lambda_{\alpha}-k_{\ell}-2c'\mathrm{i}} = - {\prod^M_{ \beta = 1} }
\frac{\Lambda_{\alpha}-\Lambda_{\beta} +2c'\mathrm{i}}{\Lambda_{\alpha}-\Lambda_{\beta}-2c'\mathrm{i}}. 
\label{BA}
\end{eqnarray}
\noindent Here $c'=c/4$, $j=1,...,N$, $\alpha=1,...,M$ and $\{ \Lambda_{\alpha} \}$ are the rapidities for the internal spin degrees of freedom. 
The quantum number $M$ is a conserved quantity obeying the relation $M=N-S^z$.
In this model, the antiferromagnetic interaction leads to an effective attraction  in the spin-singlet channel so  that  the singlet bosonic pairs comprise a spin singlet 
ground state.  
In the thermodynamic limit $N,L\rightarrow\infty$, the sets of
solutions $\{k_{j}\}$ and $\{\Lambda_{\alpha}\}$ of the BA equations (\ref{BA})
take a certain form, where the 
$k_{j}$'s and $\Lambda_{\alpha}$'s can form complex pairs
$k_{j}=\lambda_{j}\pm\mathrm{i}c'$ and
$\Lambda_{j}=\lambda_{j}\pm\mathrm{i}c'$ where $\lambda_{j}$ is
real. Notice that each pair of $k_{j}$'s share the same real part as
a corresponding pair of $\Lambda_{j}$'s. The bound states are associated
with a pair of $|F=1,m_F=\pm1\rangle $ bosons or two $|F=1,m_F=0\rangle $ bosons.
 In addition to that,
we also have real $k_{j}$'s and $\Lambda$ strings of the form
$\Lambda_{\alpha}^{n,j}=\Lambda_{\alpha}^{n}+\mathrm{i}(n+1-2j)c'$,
$j=1,\ldots,n$  describing spin wave bound states. 

At finite temperatures, the physical states become degenerate.  The equilibrium state can be obtained by the condition of minimizing the Gibbs free energy 
$G = E + E_z -\mu N - TS$, where $\mu$ is the chemical potential and $S$ the entropy, see Yang and Yang's   grand  canonical description \cite{Yang-Yang} of the BA equations for 
the integrable Bose gas.  The Zeeman	energy $E_Z = -HS_z$ and entropy $S$ 
are given in terms of the densities of charge bound states and spin-strings described above which are subject to the BA equations (\ref{BA}). 
Minimizing the Gibbs free energy leads to a set of coupled non-linear integral equations, i.e., the TBA equations (see \cite{Lee} for details)
\begin{eqnarray}
\varepsilon_1(k) &=& k^2-\mu-H-Ta_4*\ln(1+e^{-\frac{\varepsilon_1(k)}{T}})\nonumber \\ 
&&+ T[a_1-a_5]*\ln(1+e^{-\frac{\varepsilon_2(k)}{T}}) \nonumber \\ 
&&- T\sum_{n=1}^{\infty}[a_{n-1}+a_{n+1}]*\ln(1+e^{-\frac{\phi_n(k)}{T} }), \nonumber \\ 
\varepsilon_2(k)&=& 2(k^2-c'^2-\mu)+T[a_1-a_5]*\ln(1+e^{-\frac{\varepsilon_1(k)}{T}})\nonumber\\
&&+ T[a_2-a_4-a_6]*\ln(1+e^{-\frac{\varepsilon_2(k)}{T} }), \nonumber\\ 
\phi_n(k)&=&n+T[a_{n-1}+a_{n+1}]*\ln(1+e^{-\frac{\varepsilon_1(k)}{T}}) \nonumber \\ 
&&+ T\sum_{n=1}^{\infty}T_{mn}*\ln(1+e^{-\frac{\phi_n(k)}{T}}).
\label{TBA}
\end{eqnarray}
Here $n=1,2,\ldots, \infty$ and  the symbol $*$  denotes the convolution $(f*g(x))=\int_{-\infty}^{\infty}f(x-x')g(x')dx'$, the functions $a_n=\frac{1}{\pi}\frac{n|c'|}{(nc')^2+x^2}$ 
and $T_{nm}$ are given in \cite{Lee}.
These  TBA equations are expressed in terms of the dressed energies $\varepsilon_1(k)$, $\varepsilon_2(k)$ and $\phi_n(k)$ for unpaired states, paired states and spin strings, respectively. 
They depend on the chemical potential $\mu$, the external field $H$ and spin fluctuations which are ferromagnetically coupled to the unpaired Fermi sea.

The pressure per unit length of the system is given by $p=p_1+p_2$ with
\begin{eqnarray}
p_1&=&\frac{T}{2\pi}\int_{-\infty}^{\infty}\ln(1+e^{-\varepsilon_1(k)/T})dk, \nonumber\\
p_2&=&\frac{T}{\pi}\int_{-\infty}^{\infty}\ln(1+e^{-\varepsilon_2(k)/T})dk,
\label{pressure}
\end{eqnarray}
corresponding  to the pressures for unpaired bosons  and spin-singlet pairs, respectively. 

\section{Phase diagram in the $\mu-H$ plane}
\label{Phase}

The ground state properties and phase diagram at zero temperature can be determined by the   dressed energy equations 

\begin{eqnarray}
\varepsilon_1(k)&=&k^2-\mu-H+a_4*\varepsilon_1(k)+[a_5-a_1]*\varepsilon_2(k), \label{TBA1_t0} \nonumber \\ 
\varepsilon_2(k)&=&2(k^2-c'^2-\mu)+[a_5-a_1]*\varepsilon_1(k) \nonumber \\ 
&&+[a_6+a_4-a_2]*\varepsilon_2(k)
\label{TBA_t0}
\end{eqnarray}
which are obtained from the TBA equations (\ref{TBA}) in the limit $T\to 0$. 
The negative part of the dressed energies $\varepsilon_a(k)$, $a=1$, $2$ for $k\leq Q_a$ 
corresponds to occupied states, while the positive part of $\varepsilon_a$ 
corresponds to unoccupied states. The integration boundaries $Q_a$ characterize the ``Fermi surfaces'' defined by $\varepsilon_a(Q_a)=0$.
In a canonical ensemble the bosonic pairs  form a spin singlet ground state when the external field is less than a lower critical field.  
In this phase, the  low energy physics can be characterized  by a spin-charge separation theory of the $U(1)$ TLL  describing the charge sector  and a $O(3)$ non-linear sigma 
model  describing the spin sector \cite{Shlyapnikov}. However,  if the  external field exceeds an upper critical field, we have solely ferromagnetic single  bosons  with  
aligned spins  along the external field. For an intermediate  magnetic field, the spin-singlet pairs  and spin-aligned bosons form a two-component TLL  with magnetization \cite{Lee}. 
 However, in realistic  experiments with cold atoms, 1D systems can be realized by tightly confining the atomic cloud in two (radial) dimensions and weakly confining it along the axial direction in a harmonic trap. Therefore, the phase diagram in the $\mu-H$ plane is essential for understanding  quantum criticality of  the trapped gas at finite temperatures. 

We may determine the phase boundaries by analysing the band fillings in the dressed energy equations (\ref{TBA_t0}).  
The $V-F$ phase boundary is established by the condition $\varepsilon_1(k) \leq 0$ 
and $\varepsilon_2(k)> 0$. Then from equation (\ref{TBA_t0}) we have $\mu_{c1}=-H$. The $V-S$ phase boundary is determined by  $\varepsilon_1(k)>0$ and $\varepsilon_2(k)\leq 0$ that  results in $\mu_{c2}=-\epsilon_{\rm b}/2$, where $\epsilon_{\rm b} =\hbar^2c^2/(16m)$ is the binding energy of the bound pair.
For convenience we shall use the dimensionless units in the study of quantum criticality of the system, i.e.,  $\tilde{\mu}\equiv\mu/\epsilon_b$ and $h=H/\epsilon_{\rm b}$. Thus the critical fields for the phase boundaries $V-F$ and $V-S$ read 
\begin{eqnarray}
\tilde{\mu}_{c1}=-h,\qquad \tilde{\mu}_{c2}=-\frac{1}{2}. 
\label{mu_exac1}
\end{eqnarray}

The ($F-M$) phase boundary is obtained by the requirement $\varepsilon_1(\pm Q_1)=0$ and $\varepsilon_2(k)\leq 0$,
yielding the  set of equations
\begin{eqnarray}
\tilde{\varepsilon}_1(x) &=& 8x^2-\tilde{\mu}_{c3}-h + \frac{1}{\pi}\int_{-\tilde{Q}_1}^{\tilde{Q}_1}\frac{\tilde{\varepsilon}_1(x')}{1+(x-x')^2}dx', \nonumber \\
\tilde{Q}_1^2 &=& \frac{\tilde{\mu}_{c3}}{8} +\frac{h}{8} -\frac{1}{8\pi}\int_{-\tilde{Q}_1}^{\tilde{Q}_1}\frac{\tilde{\varepsilon}_1(x')}{1+(\tilde{Q}_1-x')^2}dx',
\label{vrep1}
\end{eqnarray}
which  give the critical field for the phase transition from a ferromagnetic phase of spin-aligned bosons into  a mixed phase of the  pairs  and unpaired bosons, 
\begin{eqnarray}
\tilde{\mu}_{c3} &=& -\frac{1}{2}+\frac{2}{5\pi}\int_{-\tilde{Q}_1}^{\tilde{Q}_1}\frac{\tilde{\varepsilon}_1(x)}{1+16x^2/25}dx, \nonumber \\
&-& \frac{2}{\pi}\int_{-\tilde{Q}_1}^{\tilde{Q}_1}\frac{\tilde{\varepsilon}_1(x)}{1+16x^2}dx.
\label{mu3_ep}
\end{eqnarray}
where  $\tilde{Q}_1=Q_1/c$ and $\tilde{\varepsilon}_1(x)$ is given by  (\ref{vrep1}).

The $S-M$ phase boundary is determined by the conditions  $\varepsilon_1(k)\leq 0$ and 
$\varepsilon_2(\pm Q_2)=0$, from which we obtain  the  set of equations, 
\begin{eqnarray}
\tilde{\varepsilon}_2(x) &=& 2\left(8x^2-\frac{1}{2}-\tilde{\mu}_{c4}\right)+\frac{1}{\pi}\int_{-\tilde{Q}_2}^{\tilde{Q}_2}\frac{\tilde{\varepsilon}_2(x')}{1+(x-x')^2}dx' \nonumber \\
&&+\frac{2}{3\pi}\int_{-\tilde{Q}_2}^{\tilde{Q}_2}\frac{\tilde{\varepsilon}_2(x')}{1+4(x-x')^2/9}dx'\nonumber \\ 
&&-\frac{2}{\pi}\int_{-\tilde{Q}_2}^{\tilde{Q}_2}\frac{\tilde{\varepsilon}_2(x')}{1+4(x-x')^2}dx',\nonumber\\ 
\tilde{Q}_2^2 &=& \frac{1}{16}+\frac{\tilde{\mu}_{c4}}{8}-\frac{1}{24\pi}\int_{-\tilde{Q}_2}^{\tilde{Q}_2}\frac{\tilde{\varepsilon}_2(x')}{1+4(\tilde{Q}_2-x')^2/9}dx'  \nonumber\\
&&- \frac{1}{16\pi}\int_{-\tilde{Q}_2}^{\tilde{Q}_2}\frac{\tilde{\varepsilon}_2(x')}{1+(\tilde{Q}_2-x')^2}dx'\nonumber \\
&&+\frac{1}{8\pi}\int_{-\tilde{Q}_2}^{\tilde{Q}_2}\frac{\tilde{\varepsilon}_2(x')}{1+4(\tilde{Q}_2-x')^2}dx'
\label{vrep2}
\end{eqnarray}
that provide the critical fields for a phase transition from the spin-singlet phase of paired bosons into  a mixed phase of  pairs  and unpaired bosons,
\begin{eqnarray}
\tilde{\mu}_{c4} &=& -h +\frac{4}{5\pi}\int_{-\tilde{Q}_2}^{\tilde{Q}_2}\frac{\tilde{\varepsilon}_2(x)}{1+16x^2/25}dx\nonumber \\
&&-\frac{4}{\pi}\int_{-\tilde{Q}_2}^{\tilde{Q}_2}\frac{\tilde{\varepsilon}_2(x)}{1+16x^2}dx. 
\label{mu4_ep}
\end{eqnarray}
Here  $\tilde{Q}_2=Q_2/c$ and $\tilde{\varepsilon}_2(x)$ is given by  (\ref{vrep2}).

In order to investigate quantum criticality of the system in the strong coupling regime, we need  closed form expressions  for  the critical fields. 
By Taylor expansion of Eqs. (\ref{mu3_ep}) and (\ref{mu4_ep}), we obtain the critical field values 
\begin{eqnarray}
\tilde{\mu}_{c3}&=& -\frac{1}{2} + \frac{8\sqrt{2}}{15\pi}\left(h-\frac{1}{2}\right)^{\frac{3}{2}} + \frac{104}{75\pi^2}\left(h-\frac{1}{2}\right)^2
\label{mu3_ap} \\
\tilde{\mu}_{c4}&=& -h + \frac{32\sqrt{2}}{15\pi}\left(\frac{1}{2}-h\right)^{\frac{3}{2}} + \frac{2912}{225\pi^2}\left(\frac{1}{2}-h\right)^2
\label{mu4_ap}
\end{eqnarray}
which are in good agreement with the numerical results obtained from  (\ref{mu3_ep}) and (\ref{mu4_ep}) in the strong coupling regime. 
These asymptotic results  (Eqs.(\ref{mu3_ap}) and (\ref{mu4_ap})) can also be obtained by converting the critical fields obtained  
in the $H-n$ plane \cite{Lee}  into the $\mu-H$ plane, where  the effective chemical potentials $\mu_1=\mu+H$, $\mu_2=\mu+\epsilon_{\rm b}/2$ for unpaired and 
paired bosons are presented explicitly in \cite{Lee}. 

In the next  section we will  derive analytical expressions for  the equation of state and universal TLL  
thermodynamics  in the physical regime where $t=T/\epsilon_{\rm b}\ll 1$, i.e.,  
for the strong coupling and low temperature regimes. 

\section{Equation of state and TLL thermodynamics}
\label{EOS}

The thermodynamics and the high precision of the equation of state of a system 
are the key informations that can be used to map out quantum critical phenomena and to make comparisons between theory and experiment. 
Recently, the equation of state of a two-component ultra-cold Fermi gas has been measured \cite{salomon,Horikoshi} using theoretical schemes \cite{Ho-Zhou}.  
Such experimental advances  provide exciting opportunities to test universal TLL and quantum critical phenomena in low dimensional many-body systems. 

\subsection{Equation of state}
The lack of analytic solutions of the TBA equations  limits the ability to make physical predictions of the model at finite temperatures. 
In fact, the thermodynamic properties of the  model at finite temperature are notoriously difficult  to extract due to the presence of the bosonic nature and the  spin-spin exchange interaction. 
Building on the method presented in \cite{Guan-Ho} and considering the physical region (strong coupling $|c|\gg 1$ and low temperatures), we find that
spin fluctuations are strongly suppressed by a strong field, i.e.,  $H\gg T $. Therefore we can analytically extract the  spin wave bound state contributions  to 
the unpaired dressed energy, see the third equation  in (\ref{TBA}). Moreover, we notice that  the convolution terms   converge rapidly   once $\varepsilon_{1,2}(k)>0$ 
 in the TBA equations. Therefore, 
we are  allowed to carry out a Taylor expansion with respect to $c$ in the kernel functions $a_n$ in  the  TBA equations at low temperatures. 
Then, integrating  by parts, we may  obtain  the dressed energies   in terms of polylogarithm functions up to order $1/|c|^3$, 
\begin{eqnarray}
\varepsilon_1(k) &\approx & \frac{\hbar^2}{2m}k^2-\mu-H-\frac{2|c|p_1}{c^2+k^2} +\frac{ 4|c|p_2}{c^2+16k^2}\nonumber \\
&&-\frac{ 20|c|p_2}{25c^2+16k^2}-\frac{T^{\frac{5}{2}}}{2\sqrt{\pi}|c|^3\left(\frac{\hbar^2}{2m}\right)^{\frac{3}{2}}}{\mathrm {Li}}_{\frac{5}{2}}\left(-e^{\frac{A^0_1}{T}}\right) \nonumber\\ 
&&+ \frac{1984 T^{\frac{5}{2}}}{125\sqrt{2\pi}|c|^3\left(\frac{\hbar^2}{2m}\right)^{\frac{3}{2}}}{\mathrm {Li}}_{\frac{5}{2}}\left(-e^{\frac{A^0_2}{T}}\right) \\ 
&& -Te^{-\frac{H}{T}-\frac{\bar{K}}{4}}\left[\left(1-\frac{2k^2}{c^2}\right)I_0\left(\frac{\bar{K}}{4}\right)+\frac{2k^2}{c^2}I_1\left(\frac{\bar{K}}{4}\right)\right],\nonumber 
\label{TBA1lowT}\\
\varepsilon_2(k) &\approx & \frac{2\hbar^2}{2m}k^2-\frac{\hbar^2}{2m}\frac{c^2}{8}-2\mu +\frac{8|c|p_1}{c^2+16k^2} \nonumber\\ 
&&-\frac{40|c|p_1}{25c^2+16k^2}+ \frac{3968 T^{\frac{5}{2}}}{125\sqrt{\pi}|c|^3\left(\frac{\hbar^2}{2m}\right)^{\frac{3}{2}}}{\mathrm {Li}}_{\frac{5}{2}}\left(-e^{\frac{A^0_1}{T}}\right)\nonumber  \\ 
&&+ \frac{2|c|p_2}{c^2+4k^2}-\frac{|c|p_2}{c^2+k^2}-\frac{6|c|p_2}{9c^2+4k^2}\nonumber\\
&&+\frac{181 T^{\frac{5}{2}}}{108\sqrt{2\pi}|c|^3\left(\frac{\hbar^2}{2m}\right)^{\frac{3}{2}}}{\mathrm {Li}}_{\frac{5}{2}}\left(-e^{\frac{A^0_2}{T}}\right).
\label{TBA2lowT}
\end{eqnarray}
Here   ${\mathrm {Li}}_s(z)=\sum_{k=1}^{\infty}z^k/k^s$ is the polylogarithm function. The terms 
$\bar{K}=8p_1/(T|c|)$  and $I_n(z)=\sum_{\gamma=0}^{\infty}\frac{(z/2)^{n+2\gamma}}{\gamma!(n+\gamma)!}$ are obtained  from the so-called ``string'' or spin wave contributions. 

Using the above dressed energies and integrating by parts, we may calculate the pressure   (\ref{pressure}) in a straightforward way, with result 
\begin{eqnarray}
p_1&\approx &-\frac{ T^{\frac{3}{2}}f_{\frac{3}{2}}^1}{\left(\frac{4\pi\hbar^2}{2m}\right)^{\frac{1}{2}}}\left[1-\frac{p_1}{|c|^3}\frac{2m}{\hbar^2} +\frac{3968p_2}{125|c|^3}\frac{2m}{\hbar^2}\right], \label{press1}\\
p_2&\approx &-\frac{ T^{\frac{3}{2}}f_{\frac{3}{2}}^2}{\left(\frac{2\pi\hbar^2}{2m}\right)^{\frac{1}{2}}}\left[1+\frac{3968p_1}{125|c|^3}\frac{2m}{\hbar^2} +\frac{181p_2}{108|c|^3}\frac{2m}{\hbar^2}\right], 
\label{press2}
\end{eqnarray}
where we have denoted $f_{s}^i={\mathrm {Li}}_{s}\left(-e^{\frac{A_i}{T}}\right)$ with $i=1,2$ and
\begin{eqnarray}
A_1 &=& \mu+H+\frac{2p_1}{|c|}-\frac{16p_2}{5|c|}+Te^{-\frac{H}{T}}e^{-\frac{\bar{K}}{4}}I_0\left(\frac{\bar{K}}{4}\right) \nonumber\\ 
&&+ \frac{T^{\frac{5}{2}}}{|c|^3\biggl(\frac{\hbar^2}{2m}\biggr)^{\frac{3}{2}}}\left[ \frac{1}{2\sqrt{\pi}}f_{\frac{5}{2}}^1-\frac{1984}{125\sqrt{2\pi}}f_{\frac{5}{2}}^2\right], \label{A1} \\
A_2 &=& \frac{\hbar^2}{2m}\frac{c^2}{8}+2\mu -\frac{32p_1}{5|c|}-\frac{p_2}{3|c|}\nonumber \\ 
&&- \frac{T^{\frac{5}{2}}}{|c|^3\biggl(\frac{\hbar^2}{2m}\biggr)^{\frac{3}{2}}}\left[ \frac{3968}{125\sqrt{\pi}}f_{\frac{5}{2}}^1+\frac{181}{108\sqrt{2\pi}}f_{\frac{5}{2}}^2\right]. \label{A2}
\end{eqnarray}
Thus an infinite number of TBA equations have been simplified to two coupled equations, making  the thermodynamics of the model analytically accessible.

The above expressions for the pressures (\ref{press1}) and (\ref{press2}) provide the precise equation of state of the system from which universal TLL thermodynamics 
and scaling functions near critical points can be further derived analytically.  We present a  high precision  equation of state  in  the later discussions of 
the singularities of thermodynamic properties  near the quantum critical point as the temperature tends to zero.  To evaluate physical properties 
we substitute  Eqs. (\ref{A1}) and (\ref{A2}) into Eq.(\ref{press1}) and (\ref{press2}). This provides two coupled equations for $p_1$ and $p_2$, which can 
be solved by iteration. We discuss the quantum criticality of the system using these pressures in the next Section.


 
\begin{figure}[t]
{{\includegraphics [width=0.99\linewidth]{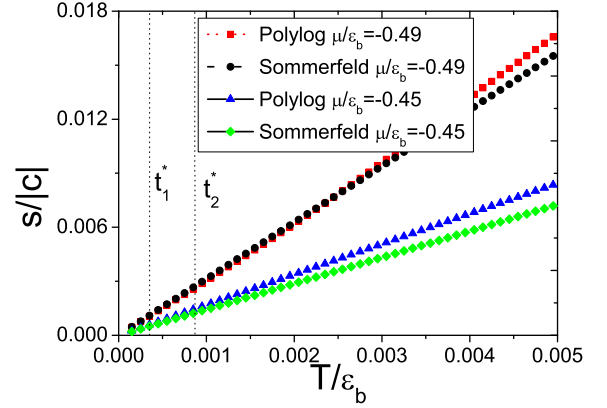}}}
\caption{(Color online) Entropy vs temperature from TLL entropy  (\ref{entropy}) and entropy calculated from the equation of state (\ref{press1} - \ref{A2}) 
for $\mu=-0.49$ and $\mu=-0.45$. The universal linear temperature dependent TLL entropy is broken down at the crossover temperatures $t^*$ which separate the TLL phase 
and the quantum critical regime, see Fig \ref{contour}.}
\label{sommer_poly} 
\end{figure}

\begin{figure}[t]
{{\includegraphics [width=0.99\linewidth]{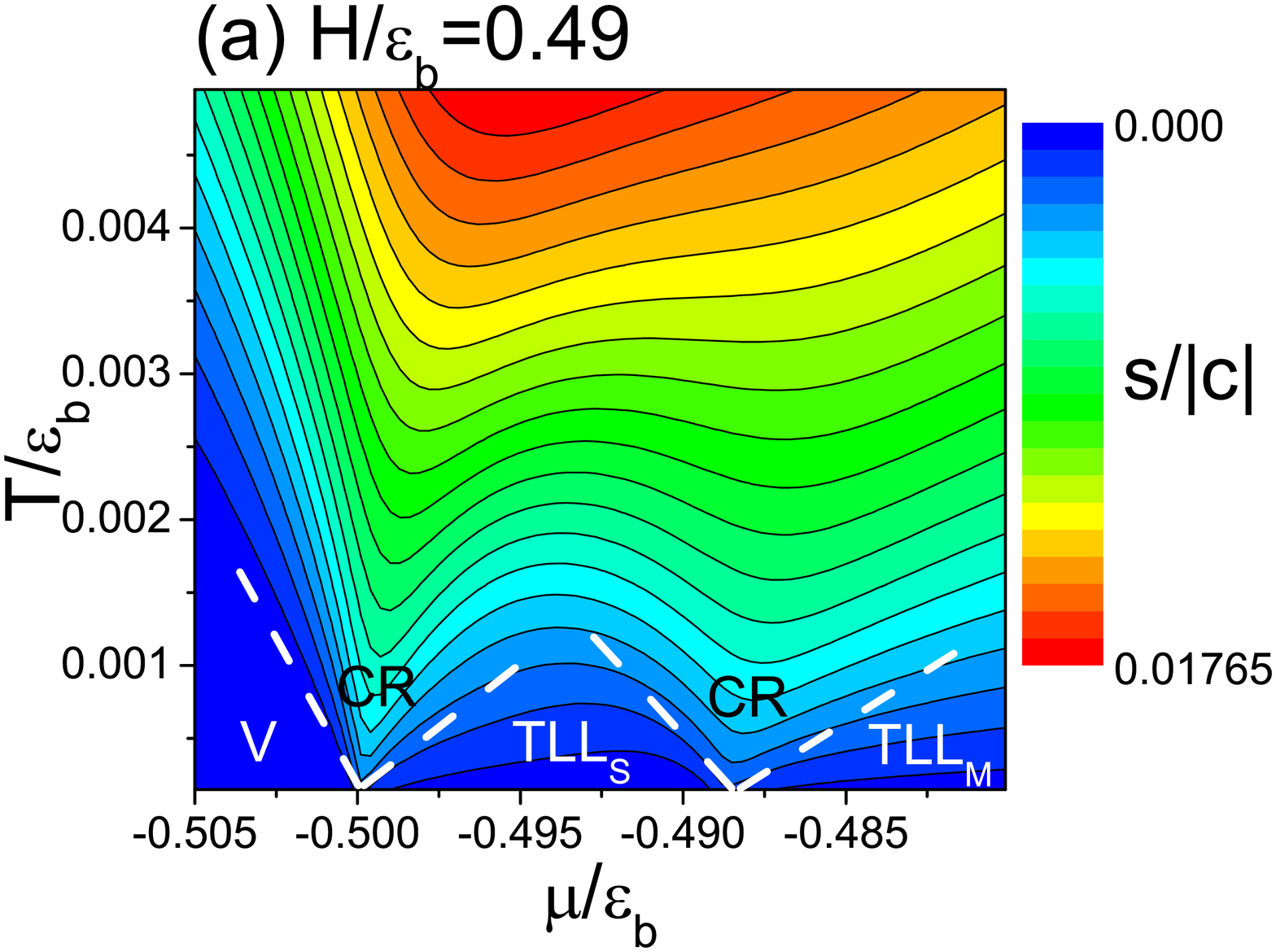}}}  \\
{{\includegraphics [width=0.99\linewidth]{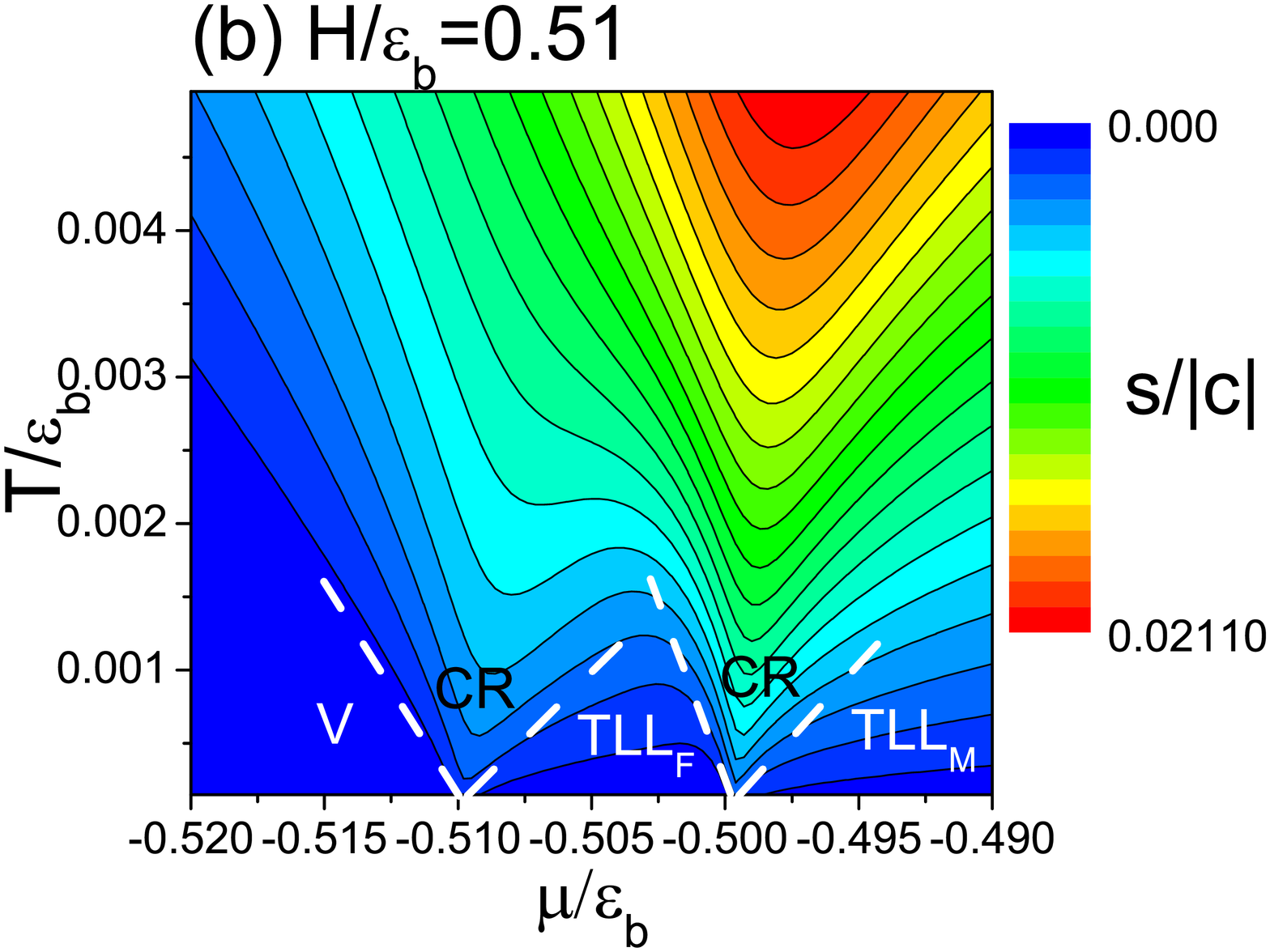}}}
\caption{(Color online) Contour plot of entropy $S$ in  the  $t-\mu$ plane from the equation of state  (\ref{press1}) and (\ref{press2}) for two values of the external magnetic 
field (a) $h=0.49$ and (b) $h=0.51$.  The dashed lines are the crossover temperatures  determined in Figure \ref{sommer_poly}.  The intersection points between two dashed lines  are the critical points in the $\mu-H$ plane, see  Figure \ref{fig1}.   (a) The dashed line separates the $TLL_S$  and  $TLL_M$  
from the quantum critical regimes.  (b) The dashed line separates the $TLL_F$  and   $TLL_M$    from the quantum critical regimes. The left-most-dashed line in (a) and (b) separates the vacuum V  from the $TLL_S$ and $TLL_F$, respectively. 
}
\label{contour} 
\end{figure}


\subsection{Universal TLL thermodynamics}
 In general, the free energy of a 1D many-body system  at  low temperatures  can be naturally attributed to low-lying excitations near the Fermi points. 
From calculations of  the   finite temperature corrections to the free energy,  one can  extract  universal TLL thermodynamics.  This low energy physics  
can also be obtained from conformal field theory  \cite{CFT}.  Here we further develop an efficient way to obtain  the universal TLL thermodynamics in the   
mixed phase of the  pairs  and unpaired bosons  from the TBA equations (\ref{TBA}).  For temperatures $T\ll 1$ and the  strong coupling regime $|c|\gg1$, the 
gapless phase is in the region $H \sim c^2\gg T$. Therefore,  we can ignore the spin wave bound state contributions in this phase. The first few terms coming from an asymptotic expansion 
in the TBA equations (\ref{TBA}) are given in terms of $1/|c|$ corrections by 
\begin{eqnarray}
\varepsilon_1(k) &\approx & \frac{\hbar^2}{2m}k^2-\mu-H-\frac{2p_1}{|c|} + \frac{16p_2}{5|c|}, \label{lowT_TBA1}\\ 
\varepsilon_2(k) &\approx & \frac{2\hbar^2}{2m}k^2-\frac{\hbar^2}{2m}\frac{c^2}{8}-2\mu+\frac{32p_1}{5|c|}+\frac{p_2}{3|c|}. \label{lowT_TBA}
\end{eqnarray}
Higher order corrections can be calculated in a straightforward manner (see the analysis in (\ref{press1}) and (\ref{press2})).  
However, they are not necessary in the present discussion.
From Eq. (\ref{pressure})  we obtain the pressures for spin-aligned single  bosons and spin-singlet pairs 
 \begin{eqnarray}
p_1 &=& \left(\frac{\hbar^2}{2m}\right)^{-1/2}\frac{1}{\pi} \int_0^{\infty}
\frac{\sqrt{\varepsilon_1^0}d\varepsilon_1^0}{1+e^{(\varepsilon_1^0 - A_1^0)/T}}, \label{p1int}\\
p_2 &=& \left(\frac{\hbar^2}{2m}\right)^{-1/2}\frac{\sqrt{2}}{\pi} \int_0^{\infty}
\frac{\sqrt{\varepsilon_2^0}d\varepsilon_2^0}{1+e^{(\varepsilon_2^0-A_2^0)/T}},
\label{p2int}
\end{eqnarray} 
\noindent where
\begin{eqnarray}
A_1^0&\approx &\mu+ H +\frac{2p_1}{|c|}-\frac{16p_2}{5|c|}, \label{a01} \\
A_2^0&\approx &2\mu+\frac{\hbar^2}{2m}\frac{c^2}{8}-\frac{32p_1}{5|c|}-\frac{p_2}{3|c|}. 
\label{a02}
\end{eqnarray}

The integrals in (\ref{p1int}) and (\ref{p2int}) can be calculated explicitly via the Sommerfeld expansion. 
We assume that there are two ``Fermi seas'', i.e., a Fermi sea of bound pairs with an effective chemical 
potential $A_2^0$ and a Fermi sea of unpaired bosons with an effective chemical  potential $A_1^0$. 
From equations (\ref{lowT_TBA1}-\ref{a02}) and the relations $n_1=\frac{\partial p}{\partial H}$, $n=n_1+2n_2=\frac{\partial p}{\partial \mu}$, 
 by a cumbersome iteration we can obtain  closed forms for  the pressures
\begin{eqnarray}
p_1 &=& \frac{2\pi^2n_1^3}{3}\left(1-\frac{6n_1}{|c|}+\frac{96n_2}{5|c|}\right) \nonumber \\
&&+ \frac{T^2}{6n_1}\left(1+\frac{2n_1}{|c|}-\frac{32n_2}{5|c|}\right), \\
p_2 &=& \frac{\pi^2n_2^3}{3}\left(1+\frac{48n_1}{5|c|}+\frac{n_2}{|c|}\right)\nonumber \\
&&+ \frac{T^2}{3n_2}\left(1-\frac{16n_1}{5|c|}-\frac{n_2}{3|c|}\right).
\label{p1p2_n1n2}
\end{eqnarray}
The Helmholtz free energy per unit length is given by $f=n\mu-p$. After a lengthy iteration,  we obtain a universal leading  temperature corrections 
to the free energy of the form 
\begin{eqnarray}
f=f_0 - \frac{\pi T^2}{6}\left(\frac{1}{v_1}+\frac{1}{v_2}\right).
\label{helm}
\end{eqnarray}
with $f_0$ the ground state already obtained in \cite{Lee}. 
Here 
\begin{equation}
v_1 = 2\pi n_1\left(1+\frac{2(32n_2-10n_1)}{5|c|}\right)
\end{equation}
and 
\begin{equation}
v_2 = \pi n_2\left(1+\frac{2(48n_1+5n_2)}{15|c|}\right)
\end{equation}
are the charge velocities for unpaired and paired bosons.
The entropy per unit length is given by  $s=-\frac{\partial f}{\partial T}$ where 
\begin{eqnarray}
s= \frac{\pi T}{3}\left(\frac{1}{v_1}+\frac{1}{v_2}\right).
\label{entropy}
\end{eqnarray}

We observe that in this gapless  phase, spin wave bound state  fluctuations  are suppressed due to a strong external field.  The
suppression of spin fluctuations leads to a universality class of a
two-component TLL  in  the mixed phase of   pairs  and unpaired bosons, which we denote by $TLL_M$. 
At low temperatures, the spin singlet phase persists  as a single component TLL (denoted by $TLL_S$)  below a crossover temperature. The fully polarised single atoms can persist  in a TLL phase (denoted by $TLL_F$) below another crossover temperature.  
However, the TLL is not appropriate to describe quantum criticality, since it does not include  proper thermal fluctuations for the quantum critical 
regime.  In general the TLL  persists  below the crossover
temperature at which  the relation of the linear temperature-dependent  entropy (or specific heat) breaks down.

In Figure \ref{sommer_poly} we present the entropy as a function of the temperature
using these two different approaches -- the polylogarithm function result (\ref{press1}-\ref{A2})  and the Sommerfeld expansion (\ref{entropy}).  The crossover 
temperature $t^{*}$ determines the boundaries between the TLL  regime and the quantum critical regime, see the contour plots of entropy in the $t-\mu$ plane  for 
two different values of the  external field in  Figure \ref{contour}. The crossover boundaries are established by the points  at which the TLL  entropy (\ref{entropy}) 
breaks down, i.e., the entropy is no longer linear temperature-dependent.  At finite temperatures, the system exhibits the characteristic 
V-shaped behaviour of quantum criticality.

\section{QUANTUM CRITICALITY}
\label{QC}

Quantum criticality describes the critical behavior near a quantum phase transition, i.e., it describes collective behavior  of a large number of interacting particles 
at temperatures sufficiently low, such that quantum mechanics plays a crucial role in determining the distinguishing characteristics \cite{sachdev}. The quantum phase transition 
occurs at absolute zero temperature as the parameters of the system  are varied. In the  critical regime, a universal and scale-invariant description of the system is expected through the 
power-law scaling of thermodynamical properties.  From the phase diagram of spin-1 bosons, see Figure \ref{fig1}, we observe that the quantum phase transition occurs as the driving 
parameters $\tilde{\mu}$ and $\tilde{h}$ cross the phase boundaries at zero temperature. Although there is no true finite temperature quantum   phase transition  in a 1D model, 
quantum criticality of a 1D many-body system is  associated with  a  universal crossover  $T^*$ that  separates the excitation spectrum  from relativistic to nonrelativistic 
dispersion \cite{Oshikawa,Erhai}. 
In the present spin-1 model of bosons, we can interpret the spin singlet phase as a TG gas of hard-core bosons with mass $2m$ and the spin-aligned  ferromagnetic phase as a TG gas  
of single atoms  with mass $m$. The mixed phase of two coupled TG gases  is made of particles with mass $m$ and $2m$.  

\subsection{Criticality driven by a chemical potential}

Quantum critical behaviour is  uniquely characterized by the critical exponents
depending only on the dimensionality and the symmetry of the excitation spectrum.
This is reflected by singularities in the thermodynamic quantities, 
such as density $n$, compressibility $\kappa=\partial n /\partial \mu$ and magnetization $M$. They can be  obtained 
from the  derivatives of the pressure $p$ with respect to $\mu$ and $H$.  In order to identify  universal scaling 
of the thermodynamic properties in the quantum critical regime, we will only take into account the first few terms in the  
equation of state (\ref{press1}) and (\ref{press2}).  To this end,  the total pressure   is simplified as 
\begin{eqnarray}
\tilde{p}=-\frac{t^{\frac{3}{2}}}{2\sqrt{2\pi}}\left(\frac{1}{2}f_{\frac{3}{2}}^1 
+ \frac{1}{\sqrt{2}}f_{\frac{3}{2}}^{2}\right),
\label{prequantum}
\end{eqnarray}
where we have defined a dimensionless pressure $\tilde{p}=p/|c|\epsilon_b$ with the potentials
\begin{eqnarray}
\tilde{A}_1 &=& \tilde{\mu}+h+2\tilde{p}_1-\frac{16\tilde{p}_2}{5}, \label{a1tilde} \\ 
\tilde{A}_2 &=& 1+2\tilde{\mu} -\frac{32\tilde{p}_1}{5}-\frac{\tilde{p}_2}{3},
\label{a2tilde}
\end{eqnarray}
and we have denoted  the function  $f_{s}^i= {\mathrm Li}_{s}\left(-e^{\frac{A_i}{T}}\right)$.

By iterating equations (\ref{prequantum}), (\ref{a1tilde}) and (\ref{a2tilde}),  we obtain the dimensionless density $\tilde{n}\equiv n/|c|$, where 
\begin{eqnarray}
\tilde{n} &=& -\frac{\sqrt{t}}{2\sqrt{2\pi}}\left(\frac{1}{2}f_{1/2}^1\Delta_1 + \sqrt{2}f_{1/2}^2\Delta_2\right),
\label{density}
\end{eqnarray}
with 
\begin{eqnarray}
\Delta_1 &=& 1 - \frac{t^{1/2}}{2\sqrt{2\pi}}f^1_{1/2} + \frac{t}{8\pi}\left(f^1_{1/2}\right)^2 + \frac{8t^{1/2}}{5\sqrt{\pi}}f^2_{1/2}  \nonumber\\
&& +\frac{2t}{15\pi}\left(f^2_{1/2}\right)^2 + \frac{12t}{25\pi\sqrt{2}}f^1_{1/2}f^2_{1/2}, \nonumber \\
\Delta_2 &=& 1 + \frac{4t^{1/2}}{5\sqrt{2\pi}}f^1_{1/2} - \frac{t}{5\pi}\left(f^1_{1/2}\right)^2 + \frac{t^{1/2}}{12\sqrt{\pi}}f^2_{1/2} \nonumber \\
&& +\frac{t}{144\pi}\left(f^2_{1/2}\right)^2 + \frac{101t}{75\pi\sqrt{2}}f^1_{1/2}f^2_{1/2}.\nonumber
\end{eqnarray}
The total density (\ref{density}) depends on the density of single atoms and the density of paired atoms.  Furthermore,  using the standard thermodynamic relations, we obtain  the magnetization $\tilde{M}\equiv M/|c|$,  
\begin{eqnarray}
\tilde{M}&=&-\frac{\sqrt{t}}{2\sqrt{2\pi}}\Biggl\{\frac{1}{2}f^1_{1/2}\biggl(1- \frac{t^{1/2}}{2\sqrt{2\pi}}f^1_{1/2} + \frac{t}{8\pi}\left(f^1_{1/2}\right)^2 \nonumber\\
&&+ \frac{32t}{25\pi\sqrt{2}}f^1_{1/2}f^2_{1/2}\biggr) + \sqrt{2}f^2_{1/2}\biggl(\frac{4t^{1/2}}{5\sqrt{2\pi}}f^1_{1/2} \nonumber \\ 
&&-\frac{t}{5\pi}\left(f^1_{1/2}\right)^2 + \frac{t}{15\pi\sqrt{2}}f^1_{1/2}f^2_{1/2}\biggr)\Biggr\}
\label{mag}
\end{eqnarray}
and the susceptibility $\tilde{\chi}\equiv\chi\epsilon_b/|c|$, 
\begin{eqnarray}
\tilde{\chi} &=& -\frac{1}{2\sqrt{2\pi t}}\Biggl\{\frac{1}{2}f^1_{-1/2}\biggl(1- \frac{3t^{1/2}}{2\sqrt{2\pi}}f^1_{1/2} + \frac{3t}{4\pi}\left(f^1_{1/2}\right)^2 \nonumber \\ 
&& + \frac{8t^{1/2}}{5\sqrt{\pi}}f^2_{1/2}+\frac{2t}{15\pi}\left(f^1_{1/2}\right)^2 + \frac{36t}{25\pi\sqrt{2}}f^1_{1/2}f^2_{1/2}\biggr) \nonumber \\
&&+ f^2_{-1/2}\frac{32t}{25\pi\sqrt{2}}\left(f^1_{1/2}\right)^2\Biggr\}.
\label{suscep}
\end{eqnarray}
By a lengthy calculation, the compressibility $\tilde{\kappa}\equiv \kappa \epsilon_b/|c|$ is given by 
\begin{eqnarray}
\tilde{\kappa} &=& -\frac{1}{2\sqrt{2\pi t}}\Biggl\{\frac{1}{2}f^1_{-1/2}\Biggl[(\Delta_1)^2- \frac{t^{1/2}}{2\sqrt{2\pi}}f^1_{1/2}\biggl(1 \nonumber \\
&-& \frac{3t^{1/2}}{2\sqrt{2\pi}}f^1_{1/2} +\frac{16t^{1/2}}{25\sqrt{\pi}}f^2_{1/2}\biggr) + \frac{2t^{1/2}}{5\sqrt{\pi}}f^2_{1/2}\biggl(1 \nonumber \\ 
&-& \frac{3t^{1/2}}{2\sqrt{2\pi}}f^1_{1/2} + \frac{197t^{1/2}}{60\sqrt{\pi}}f^2_{1/2} \biggr)\Biggr]+2\sqrt{2}f^2_{-1/2}\Biggl[(\Delta_2)^2 \nonumber \\  
&+& \frac{4t^{1/2}}{5\sqrt{2\pi}}f^1_{1/2}\left(1+\frac{11t^{1/2}}{10\sqrt{2\pi}}f^1_{1/2}+\frac{t^{1/2}}{4\sqrt{\pi}}f^2_{1/2}\right) \nonumber \\ 
&+& \frac{t^{1/2}}{12\sqrt{\pi}}f^2_{1/2}\left(1+ \frac{424t^{1/2}}{25\sqrt{2\pi}}f^1_{1/2} + \frac{t^{1/2}}{4\sqrt{\pi}}f^2_{1/2} \right)\Biggr]\Biggr\}.
\label{comp}
\end{eqnarray}

These thermodynamic properties pave a way to extract  universal scaling functions in the vicinity of the critical points $\tilde{\mu}_c$.
Following the procedure  discussed in \cite{Guan-Ho},  quantum criticality of these thermodynamic quantities can be obtained  in the limit $T\rightarrow0$ and $T>|\tilde{\mu}-\tilde{\mu}_c|$ across each of the phase boundaries, with the results 
\begin{eqnarray}
(V - F)\,\, \left\{  \begin{array}{l}
                        \tilde{n}\simeq-\frac{\sqrt{t}}{4\sqrt{2\pi}}{\mathrm{Li}}_{\frac{1}{2}}\left(-e^{\frac{\tilde{\mu}-\tilde{\mu}_{c1}}{t}}\right), \\
 \tilde{M}\simeq-\frac{\sqrt{t}}{4\sqrt{2\pi}}{\mathrm{Li}}_{\frac{1}{2}}\left(-e^{\frac{\tilde{\mu}-\tilde{\mu}_{c1}}{t}}\right),
                       \end{array}\right.
\label{vacuumferro}
\end{eqnarray}
\begin{eqnarray}
(V - S)\,\, \left\{  \begin{array}{l}
                        \tilde{n}\simeq-\frac{\sqrt{t}}{2\sqrt{\pi}}{\mathrm{Li}}_{\frac{1}{2}}\left(-e^{\frac{2(\tilde{\mu}-\tilde{\mu}_{c2})}{t}}\right), \\
 \tilde{M}\simeq\frac{2t}{5\pi\sqrt{2}}{\mathrm{Li}}_{\frac{1}{2}}\left(-e^{\frac{2(\tilde{\mu}-\tilde{\mu}_{c2})}{t}}\right)\sim 0, 
                       \end{array}\right.
\label{vacuumsinglet}
\end{eqnarray}
\begin{eqnarray}
(F - M)\,\,\left\{  \begin{array}{l}
                        \tilde{n}\simeq \tilde{n}_{03}-\lambda_1\sqrt{t}{\mathrm{Li}}_{\frac{1}{2}}\left(-e^{\frac{2(\tilde{\mu}-\tilde{\mu}_{c3})}{t}}\right), \\
 \tilde{M}\simeq \tilde{M}_{03}+ \lambda_2\sqrt{t}{\mathrm{Li}}_{\frac{1}{2}}\left(-e^{\frac{2(\tilde{\mu}-\tilde{\mu}_{c3})}{t}}\right),
                       \end{array}\right. 
\label{ferromp}
\end{eqnarray}
\begin{eqnarray}
(S - M)\,\,\left\{  \begin{array}{l}
                        \tilde{n}\simeq \tilde{n}_{04}-\lambda_3\sqrt{t}{\mathrm{Li}}_{\frac{1}{2}}\left(-e^{\frac{\tilde{\mu}-\tilde{\mu}_{c4}}{t}}\right),  \\
 \tilde{M}\simeq - \lambda_4\sqrt{t}{\mathrm{Li}}_{\frac{1}{2}}\left(-e^{\frac{\tilde{\mu}-\tilde{\mu}_{c4}}{t}}\right).
                       \end{array}\right. 
\label{singletmp}
\end{eqnarray}
Here  $\tilde{M}_{03}=\tilde{n}_{03}$, $\tilde{n}_{04}$, $\lambda_i$, with $i=1 \ldots 4$, $a$ and 
$b$ are constants, independent of $\tilde{\mu}$ and $t$. They  are given explicitly by
\begin{eqnarray}
\tilde{n}_{03}&=&\frac{\sqrt{a}}{2\pi\sqrt{2}}\left(1+\frac{\sqrt{a}}{\pi\sqrt{2}}+\frac{a}{2\pi^2}\right),\nonumber \\
\tilde{n}_{04}&=&\frac{\sqrt{b}}{\pi}\left(1-\frac{\sqrt{b}}{6\pi}+\frac{b}{36\pi^2}\right),\nonumber \\
\lambda_{1}&=&\frac{1}{2\sqrt{\pi}}\left(1-\frac{16\sqrt{a}}{5\pi\sqrt{2}}-\frac{8a}{25\pi^2}\right),\nonumber \\
\lambda_{2}&=&\frac{4\sqrt{a}}{5\sqrt{2}\pi^{3/2}}\left(1-\frac{3\sqrt{a}}{5\pi\sqrt{2}}\right),\nonumber \\
\lambda_{3}&=&\frac{1}{4\sqrt{2\pi}}\left(1-\frac{32\sqrt{b}}{5\pi}+\frac{848b}{75\pi^2}\right),\nonumber  \\
\lambda_4&=&\frac{1}{4\sqrt{2\pi}}\left(1-\frac{16\sqrt{b}}{5\pi}+\frac{8b}{15\pi^2}\right), 
\label{const}
\end{eqnarray}
with 
\begin{eqnarray}
a&=& \left(h-\frac{1}{2}\right)\left(1 + \frac{13\sqrt{2}}{15\pi}\sqrt{h-\frac{1}{2}}\right),\nonumber\\
b&=& 2\left(\frac{1}{2}-h\right)\left(1 + \frac{91\sqrt{2}}{45\pi}\sqrt{\frac{1}{2}-h}\right).\nonumber
\end{eqnarray}

In the above equations $\tilde{n}_{03}$ and $\tilde{n}_{04}$ are the background densities near the critical points $\mu_3$ and $\mu_4$, respectively. 
At quantum criticality, the   above densities can be cast into a universal scaling form (see  \cite{sachdev,Fisher,Zhou-Ho}),  e.g., 
\begin{eqnarray}
n(\mu,T)=n_0+T^{\frac{d}{z}+1-\frac{1}{\nu z}}{\cal G}\left(\frac{\mu-\mu_c}{T^{\frac{1}{\nu z}}}\right).
\label{uni_scal_dens}
\end{eqnarray}
Here the dimensionality $d=1$ and 
the scaling function ${\cal G}(x)= \lambda  {\mathrm {Li}}_{\frac{1}{2}}(x)$ with a  constant $\lambda$. 
Consequently the dynamical critical exponent $z=2$ and  the correlation length exponent $\nu =1/2$ can be read off from the universal 
scaling form (\ref{uni_scal_dens}).  We observe that the spin-1 Bose gas belongs to the same universality class  as  spin-1/2 attractive fermions \cite{Guan-Ho} due to the hard-core nature of the two coupled Tonks-Girardeau gases. 

\begin{figure}[t]
{{\includegraphics [width=0.99\linewidth]{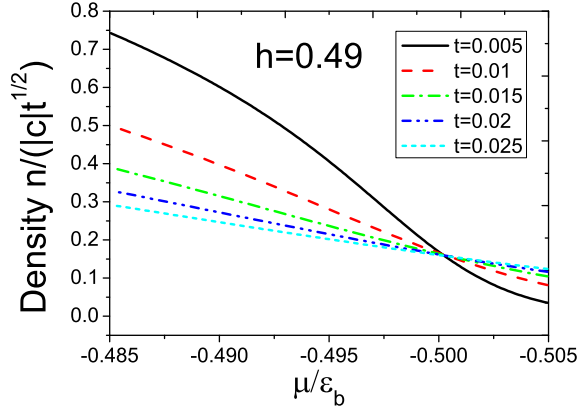}}} \\ 
{{\includegraphics [width=0.99\linewidth]{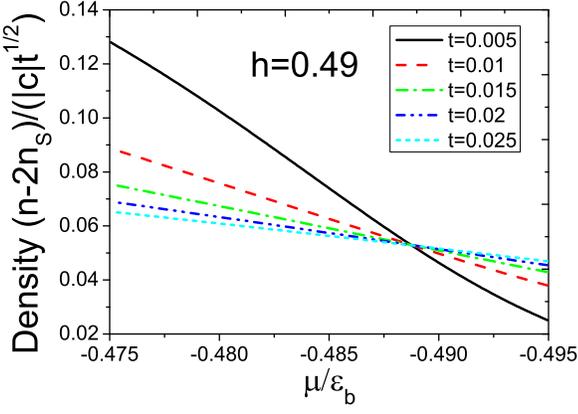}}}
\caption{(Color online) ``Scaled density'' vs chemical potential for $h=0.49$ at different 
temperatures $T/\epsilon_b=0.005$, $0.01$, $0.015$, $0.02$ and $0.025$. 
The  density curves at different temperatures  intersect  at the critical points. This feature can be used to  map out  the phase  boundary $V - S$ at $\tilde{\mu}_{c2}=-0.5$  (Eq. (\ref{mu_exac1}))  in upper panel and 
map out  the phase boundary $S - M$ at $\tilde{\mu}_{c4}\approx-0.489$ (Eq. (\ref{mu4_ap})) in the lower panel.}
\label{denityH049} 
\end{figure}

\begin{figure}[t]
{{\includegraphics [width=0.99\linewidth]{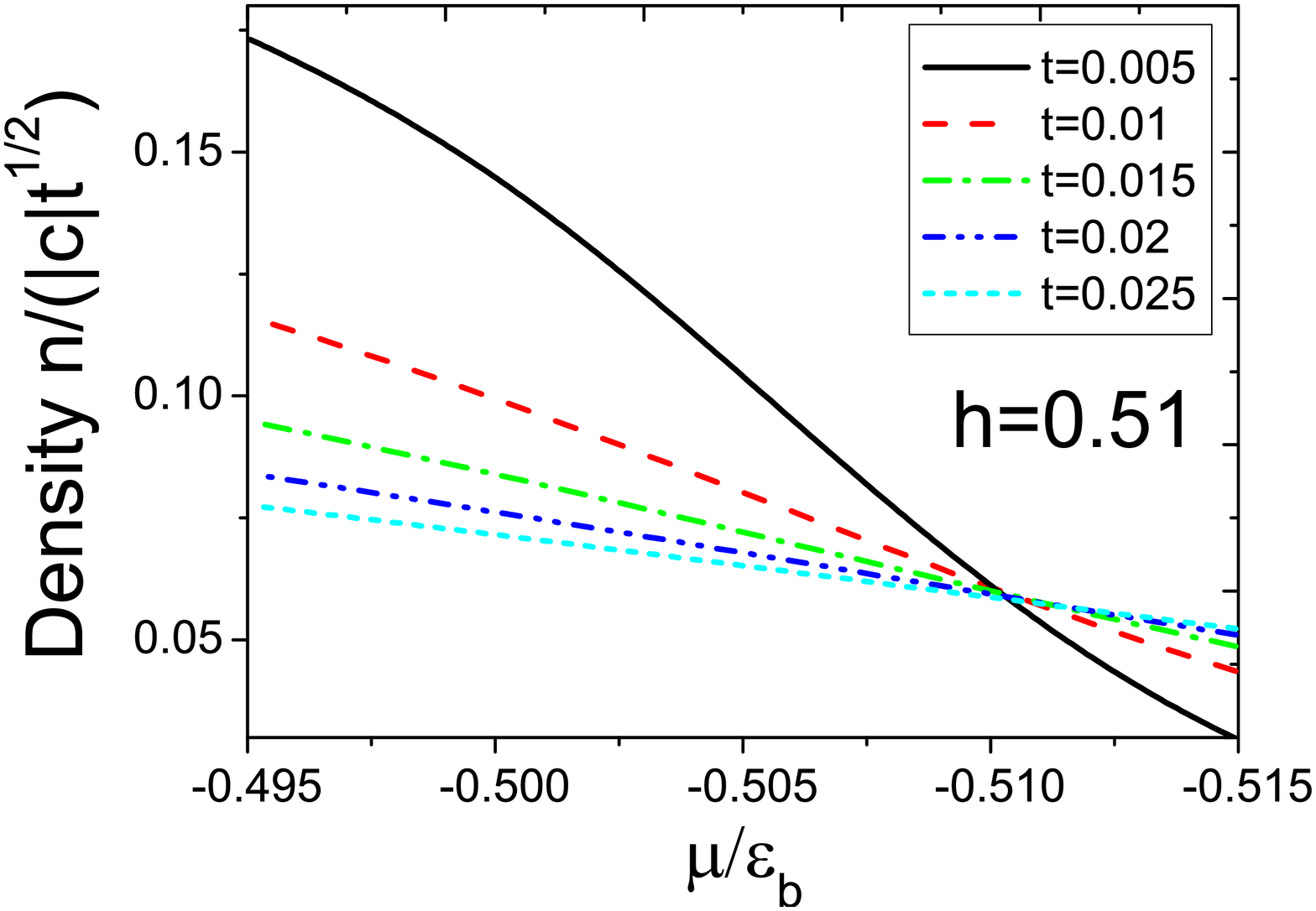}}}\\ 
{{\includegraphics [width=0.99\linewidth]{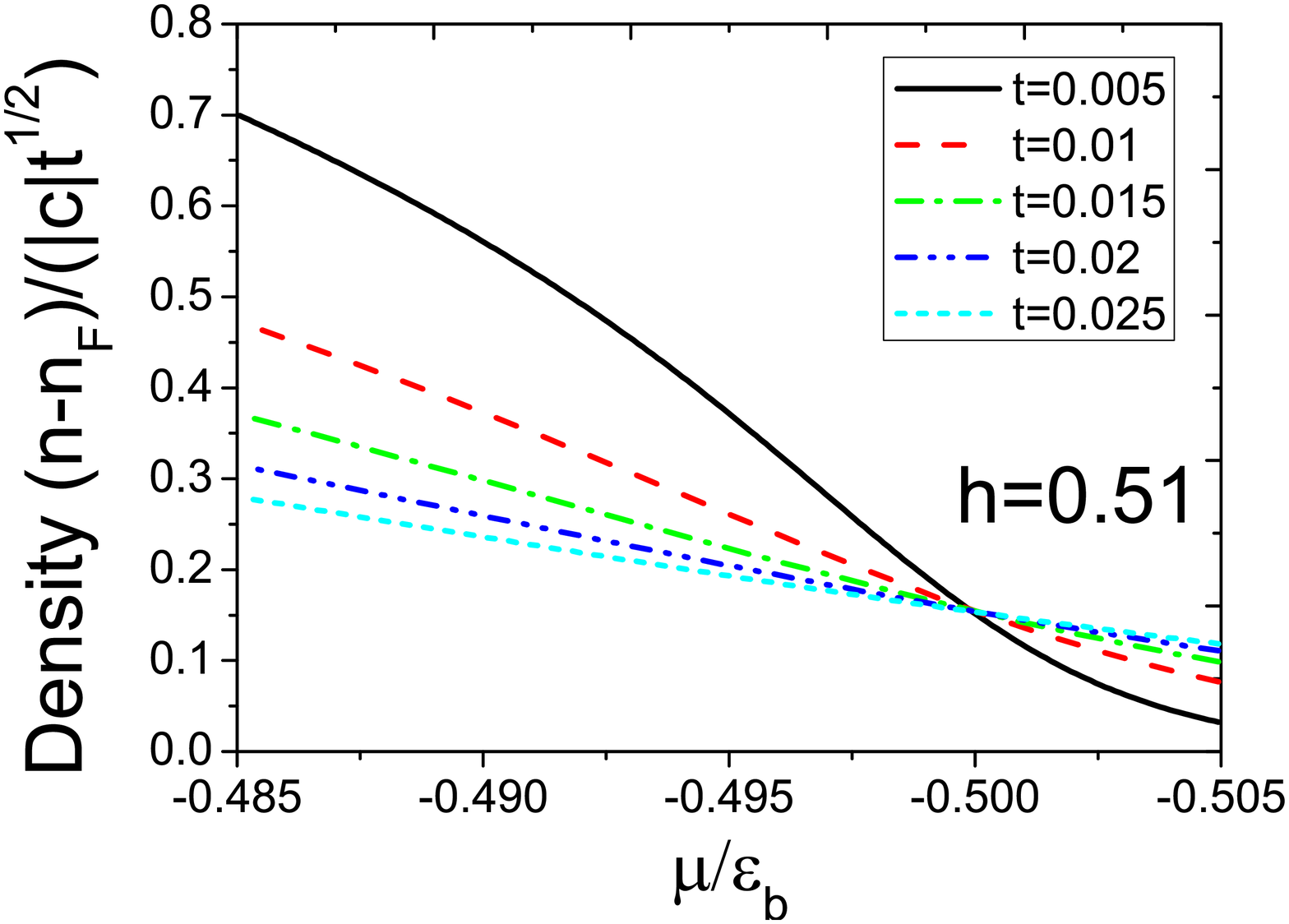}}}
\caption{(Color online) ``Scaled density'' vs chemical potential for $h=0.51$ 
at different temperatures $T/\epsilon_b=0.005$, $0.01$, $0.015$, $0.02$ 
and $0.025$. The  density curves at different temperatures  intersect at the critical points.  This feature can be used to  map out  the phase  boundary $(V - F)$ at $\tilde{\mu}_{c1}=-h$  (Eq. (\ref{mu_exac1}))  in upper panel and map out  the phase boundary $(F - M)$ at $\tilde{\mu}_{c3}\approx-0.499$ (Eq. (\ref{mu3_ap})) in the lower panel.}
\label{denityH051} 
\end{figure}

In Figures \ref{denityH049} and \ref{denityH051} we plot the ``scaled density"  $T^{-(\frac{d}{z}+1-\frac{1}{\nu z})}(n(\mu,T)-n_0)$ 
versus $\tilde{\mu}$ for different values of the temperature near the critical points $\tilde{\mu}_{c1}$, $\tilde{\mu}_{c2}$, $\tilde{\mu}_{c3}$ and $\tilde{\mu}_{c4}$.
We observe that after an appropriate  subtraction of the background density  all curves at different temperatures 
intersect at the  critical points, which is the hallmark of criticality.   In the regime of low polarization, i.e.  $P < P_c$,  the true phase transitions   from the
vacuum into the spin-singlet paired  phase and from the pure paired phase into the mixture of spin-singlet pairs and  spin-aligned bosons  occur as   the chemical potential passes 
the lower critical point $\tilde{\mu}_{c2}$ (\ref{mu_exac1}) and the upper critical point  $\tilde{\mu }_{c4}$ (\ref{mu4_ap}), respectively, see Figure \ref{denityH049}.

For large polarization, i.e., $P>P_c$, the phase transitions  from vacuum into the  ferromagnetic spin-aligned  boson phase and from the spin-aligned boson phase into 
the mixture of spin-singlet pairs and  spin-aligned bosons  occur as   the chemical potential varies across  the lower critical point $\tilde{\mu}_{c1}$ (\ref{mu_exac1}) and the upper critical point  $\tilde{\mu }_{c3}$ (\ref{mu3_ap}), respectively, see Figure \ref{denityH051}.  The universal scaling behaviour and the zero temperature phase diagram can be identified from the finite temperature density profiles of the trapped gas where the local chemical potentials are replaced by the harmonic trapping potential. 

Furthermore, we mention that similar calculations of the scaling function  can be constructed for the compressibility across all phase boundaries,
\begin{eqnarray}
&(V - F)&\,\, \tilde{\kappa}\simeq-\frac{1}{4\sqrt{2\pi t}}{\mathrm {Li}}_{-\frac{1}{2}}\left(-e^{\frac{\tilde{\mu}-\tilde{\mu}_{c1}}{t}}\right),
\nonumber \\
&(V - S)&\,\, \tilde{\kappa}\simeq-\frac{1}{\sqrt{\pi t}}{\mathrm {Li}}_{-\frac{1}{2}}\left(-e^{\frac{2(\tilde{\mu}-\tilde{\mu}_{c2})}{t}}\right),
\nonumber \\
&(F - M)&\,\, \tilde{\kappa}\simeq \tilde{\kappa}_{03}-\frac{\lambda_5}{\sqrt{t}}{\mathrm {Li}}_{-\frac{1}{2}}\left(-e^{\frac{2(\tilde{\mu}-\tilde{\mu}_{c3})}{t}}\right),
\nonumber \\
&(S - M)&\,\, \tilde{\kappa}\simeq \tilde{\kappa}_{04}-\frac{\lambda_6}{\sqrt{t}}{\mathrm {Li}}_{-\frac{1}{2}}\left(-e^{\frac{\tilde{\mu}-\tilde{\mu}_{c4}}{t}}\right).
\label{singletmpkappa}
\end{eqnarray}
Here $\tilde{\kappa}_{03}$, $\tilde{\kappa}_{04}$ are the background compressibility in the vicinities of the critical points $\tilde{\mu}_{c3}$  (Eq. (\ref{mu3_ap})) and $\tilde{\mu}_{c4}$ (Eq. (\ref{mu4_ap})), whereas   $\lambda_5$ and $\lambda_6$ are temperature-independent constants 
\begin{eqnarray}
\tilde{\kappa}_{03}&=&\frac{1}{4\pi\sqrt{2a}}\left(1+\frac{3\sqrt{a}}{\pi\sqrt{2}}+\frac{3a}{\pi^2}\right),\\
\tilde{\kappa}_{04}&=&\frac{1}{\pi\sqrt{b}}\left(1-\frac{\sqrt{b}}{2\pi}+\frac{b}{6\pi^2}\right),\\
\lambda_{5}&=&\frac{1}{\sqrt{\pi}}\left(1-\frac{24\sqrt{a}}{5\pi\sqrt{2}}+\frac{36a}{25\pi^2}\right),\\
\lambda_{6}&=&\frac{1}{4\sqrt{2\pi}}\left(1-\frac{36\sqrt{b}}{5\pi}+\frac{414b}{25\pi^2}\right).
\label{const2}
\end{eqnarray}
Again, the critical exponents $z=2$ and $\nu=1/2$ can be read off the universal scaling function ${\cal F}(x)= \lambda {\mathrm {Li}}_{-\frac{1}{2}}(x)$ in the universal form 
\begin{eqnarray}
\kappa(\mu,T)=\kappa_0+T^{\frac{d}{z}+1-\frac{2}{\nu z}}{\cal F}\left(\frac{\mu-\mu_c}{T^{\frac{1}{\nu z}}}\right). 
\label{uni_scal_kappa}
\end{eqnarray}

\subsection{Criticality driven by a magnetic field}

The quantum phase transitions driven by a magnetic field are particularly interesting. In the phase diagram  Figure \ref{fig1}, for fixed 
chemical potential we  can vary the external field $H$ to pass the phase boundaries   ($S-M$) and  ($M-F$). At finite temperatures, the 
three zero temperature  quantum phases, i.e.,  the phase of singlet pairs,  ferromagnetic phase of spin-aligned  atoms and the mixed phase 
of pairs and single atoms,  become the relativistic  TLL of bound pair ($TLL_S$), TLL of  single atoms ($TLL_F$) and a two-component TLL  ($TLL_M$) of 
paired and single atoms, respectively.
We  obtain the critical fields by converting the critical fields   Eq. (\ref{mu4_ap}) and  Eq. (\ref{mu3_ap})
\begin{eqnarray}
h_{c1}&=& -\tilde{\mu} + \frac{32\sqrt{2}}{15\pi}\left(\tilde{\mu}+\frac{1}{2}\right)^{\frac{3}{2}} - \frac{32}{45\pi^2}\left(\tilde{\mu}+\frac{1}{2}\right)^2,\label{hc1_ap} \\
h_{c2}&=& -\tilde{\mu} + \frac{1}{2}\left(\frac{15\pi}{4}\right)^{\frac{2}{3}}\left(\tilde{\mu}+\frac{1}{2}\right)^{\frac{2}{3}} - \frac{5}{8}\left(\tilde{\mu}+\frac{1}{2}\right).
\label{hc2_ap}
\end{eqnarray}

\begin{figure}[t]
{\includegraphics [width=0.99\linewidth]{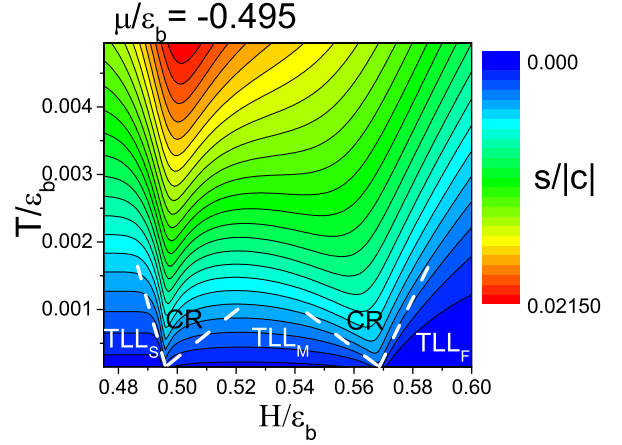}}
\caption{(Color online) Contour plot of the entropy $S$ vs the external field $H$ for  fixed  chemical potential $\tilde{\mu}=-0.495$  in the $T-H$ plane. 
The dashed lines are determined by comparing the result from the equation of state (\ref{press1}-\ref{press2}) and the TLL entropy (\ref{entropy}). 
The crossover temperatures  separates   the quantum critical regimes from the TLL phases. }
\label{contourH} 
\end{figure}

These critical fields  and the crossover temperatures  can be observed in the contour plot of  the entropy in the $T-H$ plane, see  Figure \ref{contourH}. 
The low energy  TLL physics  breaks down at the crossover temperature (dashed lines)  where the dispersion of either bound pairs or unpaired single atoms becomes nonrelativistic. 
In particular, in the vicinity of the quantum critical points $h_{c1}$  and $h_{c2}$, the system exhibits two different quantum critical regimes.
From equations (\ref{density}),(\ref{mag}) and (\ref{comp}),  we  find the scaling function 
in the critical regime near the critical point  (\ref{hc1_ap})

\begin{eqnarray}
(S - M)\,\,\left\{  \begin{array}{l}
                        \tilde{n}\simeq \tilde{n}_{05}-\lambda_7\sqrt{t}{\mathrm {Li}}_{\frac{1}{2}}\left(-e^{\frac{h-h_{c1}}{t}}\right), \\
 \tilde{M}\simeq - \lambda_8\sqrt{t}{\mathrm {Li}}_{\frac{1}{2}}\left(-e^{\frac{h-h_{c1}}{t}}\right),\\
\tilde{\kappa}\simeq \tilde{\kappa}_{05}-\frac{\lambda_9}{\sqrt{t}}{\mathrm {Li}}_{-\frac{1}{2}}\left(-e^{\frac{h-h_{c1}}{t}}\right),
                       \end{array}\right. 
\label{singletmp_H}
\end{eqnarray}
where the constants are given by 
\begin{eqnarray}
\tilde{n}_{05}&=&\frac{\sqrt{d}}{\pi}\left(1-\frac{\sqrt{d}}{6\pi}+\frac{d}{36\pi^2}\right),\nonumber\\
\tilde{\kappa}_{05}&=&\frac{1}{\pi\sqrt{d}}\left(1-\frac{\sqrt{d}}{2\pi}+\frac{d}{6\pi^2}\right),\nonumber\\
\lambda_{7}&=&\frac{1}{4\sqrt{2\pi}}\left(1-\frac{32\sqrt{d}}{5\pi}+\frac{848d}{75\pi^2}\right),\nonumber\\
\lambda_{8}&=&\frac{1}{4\sqrt{2\pi}}\left(1-\frac{16\sqrt{d}}{5\pi}+\frac{8d}{15\pi^2}\right),\nonumber \\
\lambda_{9}&=&\frac{1}{4\sqrt{2\pi}}\left(1-\frac{36\sqrt{d}}{5\pi}+\frac{414d}{25\pi^2}\right),\nonumber 
\end{eqnarray}
with
\begin{eqnarray}
 d&=& 2\left(\tilde{\mu}+\frac{1}{2}\right)\left(1 - \frac{\sqrt{2}}{9\pi}\sqrt{\tilde{\mu}+\frac{1}{2}}\right),\nonumber\\
e&=& \frac{1}{2}\left(\frac{15\pi}{4}\right)^{\frac{2}{3}}\left(\tilde{\mu}+\frac{1}{2}\right)^{\frac{2}{3}}.\nonumber
\end{eqnarray}

In the vicinity of the quantum critical points  $h_{c2}$, we obtain the scaling forms
\begin{eqnarray}
(F - M)\,\,\left\{  \begin{array}{l}
                        \tilde{n}\simeq \tilde{n}_{06}-\lambda_{10}\sqrt{t}{\mathrm {Li}}_{\frac{1}{2}}\left(-e^{\frac{\alpha(h-h_{c2})}{t}}\right), \\
 \tilde{M}\simeq \tilde{M}_{06} + \lambda_{11}\sqrt{t} {\mathrm {Li}}_{\frac{1}{2}}\left(-e^{\frac{\alpha(h-h_{c2})}{t}}\right),\\
\tilde{\kappa}\simeq \tilde{\kappa}_{06}-\frac{\lambda_{12}}{\sqrt{t}} {\mathrm {Li}}_{-\frac{1}{2}}\left(-e^{\frac{\alpha(h-h_{c2})}{t}}\right),
                       \end{array}\right. 
\label{ferromp_H}
\end{eqnarray}
where 
\begin{eqnarray}
\tilde{n}_{06}&=&\frac{\sqrt{e}}{2\pi\sqrt{2}}\left(1+\frac{\sqrt{e}}{\pi\sqrt{2}}+\frac{e}{2\pi^2}\right),\nonumber \\
\tilde{\kappa}_{06}&=&\frac{1}{4\pi\sqrt{2e}}\left(1+\frac{3\sqrt{e}}{\pi\sqrt{2}}+\frac{3e}{\pi^2}\right).\nonumber\\
\lambda_{10}&=&\frac{1}{2\sqrt{\pi}}\left(1-\frac{16\sqrt{e}}{5\pi\sqrt{2}}-\frac{8e}{25\pi^2}\right),\nonumber \\
\lambda_{11}&=&\frac{4\sqrt{e}}{5\sqrt{2}\pi^{3/2}}\left(1-\frac{3\sqrt{e}}{5\pi\sqrt{2}}\right),\nonumber \\
\lambda_{12}&=&\frac{1}{\sqrt{\pi}}\left(1-\frac{24\sqrt{e}}{5\pi\sqrt{2}}+\frac{36e}{25\pi^2}\right)
\end{eqnarray}
with 
\begin{eqnarray}
\alpha=-\frac{8}{5\pi}\left(\frac{15\pi}{4}\right)^{\frac{1}{3}}\left(\tilde{\mu}+\frac{1}{2}\right)^{\frac{1}{3}}.
\end{eqnarray}
In this case, the background density 
$\tilde{M}_{06}=\tilde{n}_{06}$ is equal to the total density at the critical point.
The density (or magnetization) and compressibility  can be recast  into the  universal scaling form

\begin{eqnarray}
M(h,T)&=&n_0+T^{\frac{d}{z}+1-\frac{1}{\nu z}}{\cal G}\left(\frac{\alpha (h-h_c)}{T^{\frac{1}{\nu z}}}\right),
\label{uni_scal_dens_h}\\
\kappa(\mu,T)&=&\kappa_0+T^{\frac{d}{z}+1-\frac{2}{\nu z}}{\cal F}\left(\frac{\alpha (h-h_c)}{T^{\frac{1}{\nu z}}}\right), 
\label{uni_scal_comp_h}
\end{eqnarray}
with  the same critical exponents as that for  quantum criticality driven by the chemical potential, i.e. the dynamical critical exponent $z=2$ and  the correlation length 
exponent $\nu =1/2$.

\begin{figure}[t]
{{\includegraphics [width=0.99\linewidth]{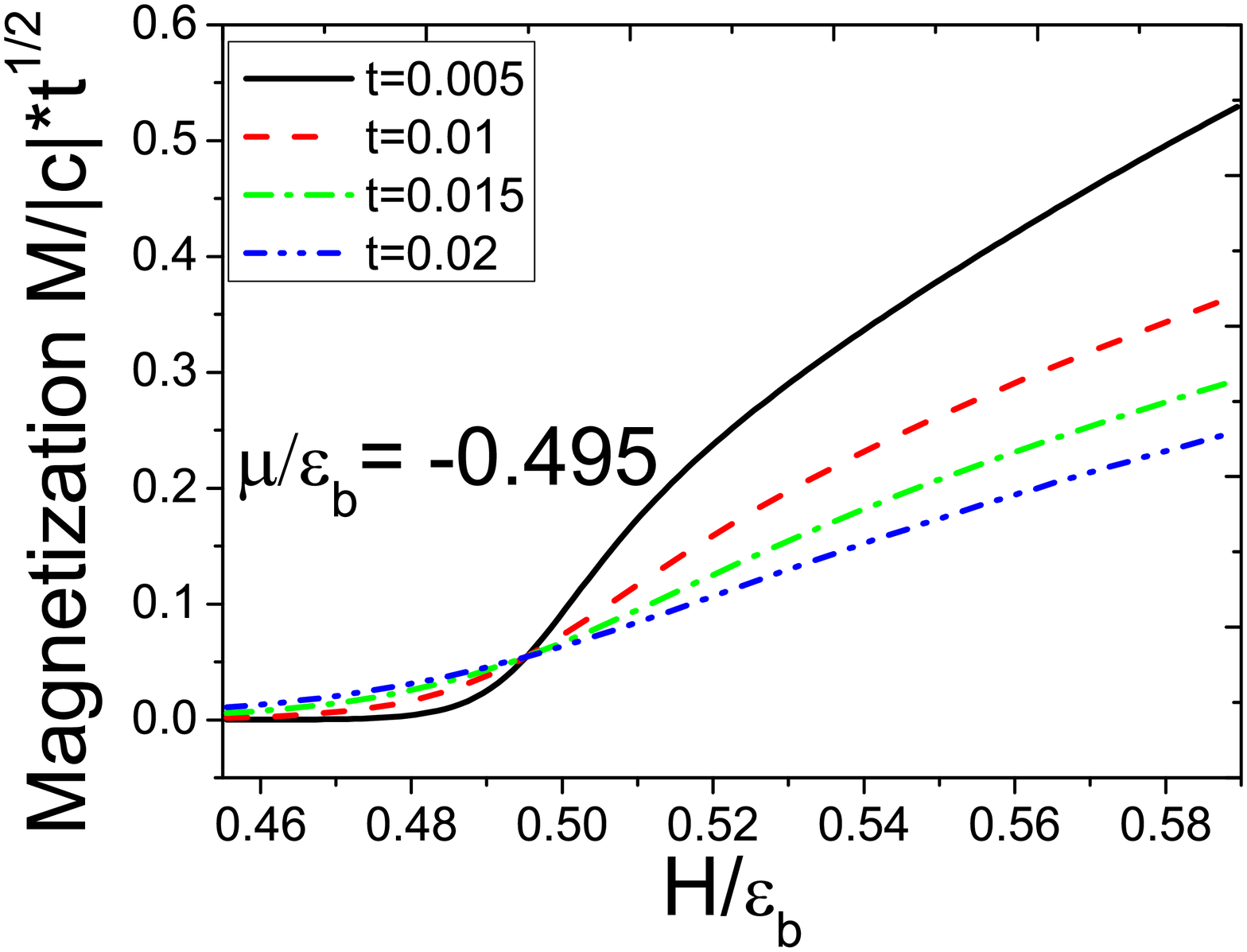}}}\\ 
{{\includegraphics [width=0.99\linewidth]{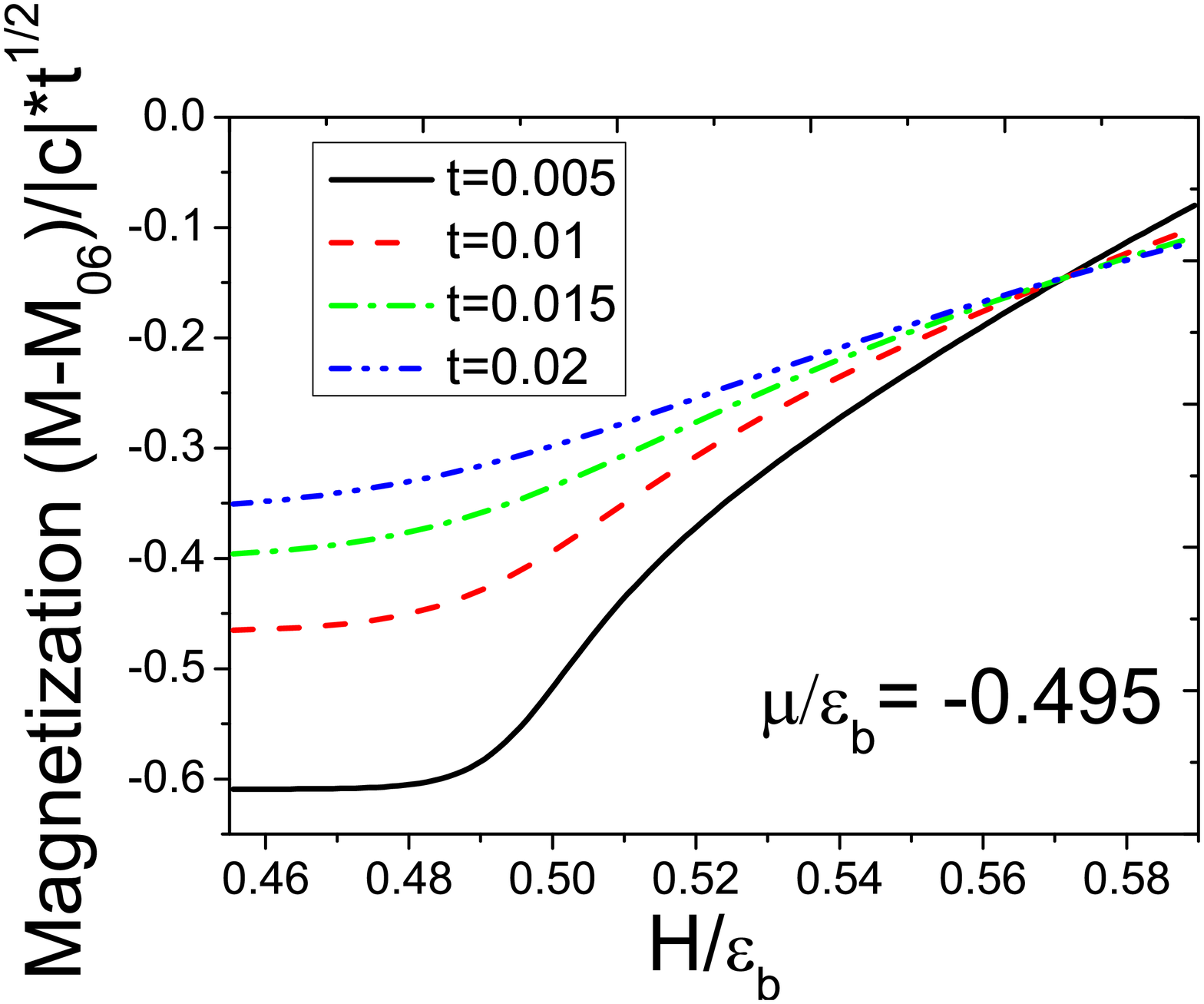}}}
\caption{(Color online) Magnetization vs external field  
for $\tilde{\mu}=-0.495$. Upper panel: the intersection point of the magnetization curves at different temperatures gives the critical external field $h_{c1}$ at 
 the boundary $S - M$. Lower panel:  after a proper subtraction of the background magnetization $\tilde{M}_{06}=n(t)$,  the intersection point of the magnetization curves at different 
temperatures gives the critical external field $h_{c2}$ at 
 the boundary $F - M$.}
\label{magH} 
\end{figure}

In Figure \ref{magH} we show the magnetization as a function of the external field
for different temperatures for a fixed chemical potential. All 
curves intersect at the  critical point $h_{c1}$ without background magnetization. However, these curves intersect at the upper critical point $h_{c2}$ with appropriate  subtraction 
of the background magnetization. We can obtain similar scaling behaviour for  the densities near the critical points  $h_{c1}$ and $h_{c2}$ like that  presented in 
Figures \ref{denityH049} and \ref{denityH051}. It turns out that magnetization can be used to map out the bulk phase diagram through the 1D  trapped gas at finite temperatures. 

\section{CONCLUSION}
\label{conclusion}

Using the TBA equations, we have studied the quantum  phase diagram, thermodynamics  and quantum critical behaviour of 
one-dimensional spin-1 bosons with strongly repulsive density-density and
antiferromagnetic spin-exchange interactions.
We have determined with high precision the equation of state from which the TLL thermodynamics, universal scaling functions and critical exponents have been obtained.  
The universality class of quantum criticality has also been discussed. 

The  scaling forms of the density, compressibility, 
magnetization and susceptibility  associated with the phase transitions driven by the chemical potential and magnetic field were rescaled  to  the universality  class of  
quantum criticality  of free fermions with  critical exponent $z=2$  and  correlation length exponent $\nu=1/2$.  
It thus  turns out that the quantum criticality of the spin-1 Bose gas belongs to the same universality class  as   spin-1/2 attractive fermions \cite{Guan-Ho} due to the 
hard-core nature of the two coupled Tonks-Girardeau gases.  We have also shown that the  quantum criticality in 1D systems  involves a universal crossover from  a TLL with 
linear dispersion to  free fermions  with a quadratic dispersion near the critical point.  These scaling forms for the thermodynamic  properties across the phase boundaries 
 illustrate the physical origin of quantum criticality in this system, where the singular part of the 
 thermodynamic properties involves a sudden change of density of state for either pairs or unpaired single atoms.  
The phase diagram, 
the TLL thermodynamics  and critical properties of the bulk system can be mapped out from the density and magnetization  profiles of the  trapped spinor  gas at finite temperatures.  
Our results open the way to further study such universal features  of 1D many-body physics in experiments with ultracold atoms.

\subsection{Acknowledgements}This work has been partially supported by the Australia Research Council. 
C. C. N. K. thanks CAPES (Coordena\c{c}\~ao de Aperfeicoamento de Pessoal de Nivel Superior) for financial support
He also thanks the Department of Theoretical Physics for their hospitality. A. F. thanks CNPq (Conselho Nacional de Desenvolvimento 
Cientifico e Tecnol\'ogico) for financial support.

\end{document}